\documentclass[10pt, a4paper, conference]{IEEEtran}

\usepackage{cite}
\usepackage{graphicx}
\usepackage[cmex10]{amsmath}
\usepackage{amssymb}
\usepackage{subfigure}
\usepackage{graphicx}
\usepackage{caption}
\usepackage{url}
\newcommand{\subparagraph}{}
\usepackage{titlesec}
\usepackage[T1]{fontenc}
\usepackage{ae,aecompl}
\usepackage{pslatex}
\usepackage{float}
\usepackage{epsfig}
\usepackage{color}
\usepackage{algorithmic}
\usepackage{todonotes}

\ifCLASSINFOpdf

\else

\fi

\begin{document}
%
\title{Proactive Dependability Framework for Smart Environment Applications}

\author{Ehsan Ullah Warriach, Tanir Ozcelebi, Johan J. Lukkien\\
Department of Mathematics and Computer Science\\
Eindhoven University of Technology\\
Eindhoven, The Netherlands\\
\{e.u.warriach, t.ozcelebi, j.j.lukkien\}@tue.nl}

\maketitle

\begin{abstract}
Smart environment applications demand novel solutions for managing quality of services, especially availability and reliability at run-time. The underlying systems are changing dynamically due to addition and removal of system components, changing execution environments, and resources depletion. Therefore, in such dynamic systems, the functionality and the performance of smart environment applications can be hampered by faults. In this paper, we follow a proactive approach to anticipate system state at runtime. We present a proactive dependability framework to prevent faults at runtime based on predictive analysis to increase availability and reliability of smart environment applications, and reduce manual user interventions.
\end{abstract}

\IEEEpeerreviewmaketitle

\section{Introduction}
A Smart Environment (SE) is a physical space enriched with embedded Information Communications Technology (ICT) and adequate software modules, and aims at creating an intelligent and reliable human-centric environment that facilitates humans to use applications efficiently. Physical components together with middleware components make up a system which provides an infrastructure/platform to build SE applications. With ever-growing complexity and dynamicity of SE systems, there is a need to ensure dynamically that a system is performing all functions according to the given specifications e.g., connectivity, and communication, and maintain sufficient level of system resources e.g., energy (battery), and memory. The adoption of SEs is hindered by the fact that there is a constant need for human (or even expert) intervention and the cost of maintenance of such systems is very high. Thus, dependable systems are required, evolving at runtime to maximize the availability and reliability of SE applications.

A SE system can support various applications. Each application can have different requirements from the underlying system. System normal operation states (from the perspective of an application) are those states that maintain the normal operation of the application according to the given specifications. An application failure is defined as the application not being able to satisfy its defined specifications~\cite{avizienis:04}. For example, in smart lighting, the application state space could be three dimensional: 1) the set of acceptable or desired light settings when users are present, 2) the set of acceptable light settings in the absence of users, and 3) the maximum delay from the time a user enters the room until the time light sources react. Obviously, the dimensions of the application state can change based on the application's ultimate goal(s). When the application does not satisfy the maximum delay requirement or provides an unacceptable light setting in the presence and in the absence of users, the application is said to be in a failure state. Otherwise it is said to be in a normal state.

\begin{figure}[htbp]
\centering
		\includegraphics[height=2.2cm,width=6.3cm]{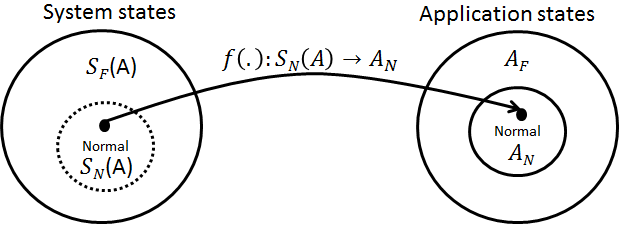}
	\caption{Mapping from the system states to the application states}
	\label{fig:boundary}
\end{figure}

Figure~\ref{fig:boundary} shows that the minimum requirements of an application define a boundary around the set of acceptable (or normal operation) states $A_{N}$ in its state space, whose dimensions are defined by application quality attributes. The set of all application failure states is $A_{F}$. $A_{F}$ and $A_{N}$ together span the entire state space of the application. The left side of Figure~\ref{fig:boundary} shows the system state space, whose dimensions are system resources and functions. A given system is said to be in a normal operation state within $S_{N}(A)$ with respect to an application $A$, if the system in its current state $S_{N}(A)$ leads to the normal operation of the application $A_{N}$. The states of the system in $S_{N}(A)$ are mapped to (give rise to) the application states in $A_{N}$. In other words, $f(.): S_{N}(A) \rightarrow A_{N}$, where $S_{N}(A)$ denotes system normal operation states for application $A$, $S_{F}(A)$ represents system failure states for application $A$, $A_{N}$ represents application normal operation states. When the system state changes from normal to failure (e.g., low Received Signal Strength Indication - RSSI below a certain threshold), this causes the application to fail, i.e., lead the application state to a state outside of $A_{N}$ and inside $A_{F}$. The boundary of $A_{N}$ is defined by the application specifications. The boundary of $S_{N}(A)$ (those system states that map onto $A_{N}$) may not be known a priori, but can be learned at runtime. As a result, learned system normal and failure states' boundaries of system resources and functions (which are not explicitly defined in the specifications of an application) will be used to predict anomalous events.


In this paper, we present a proactive dependability framework to prevent faults at the system layer regardless of the knowledge of applications. As a result, the SE system will be able to perform its functions with given specifications at hardware and software levels in the presence of an abnormal situation e.g., network node failure, communication error, interference, and battery depletion. We aim for proactive adaptation instead of reactive adaptation mechanisms~\cite{warriach:14b}. Proactive adaptation refers to the case in which the need for adaptation is anticipated, and thus action can be taken to prevent faults that are predicted by the monitoring of the system resources and functions at runtime. We have decomposed the problem of self-learning into two loosely coupled learning problems, to realize the application independent solutions for the proposed dependability framework. First, the system needs to learn at runtime the boundaries of normal and failure states of system resources and functions while running an application. Secondly, the system needs to learn effective adaptation policies for maintaining normal system states.


\section{Proactive Dependability Framework}
\begin{figure}[htbp]
\centering
		\includegraphics[height=3.9cm,width=8.6cm]{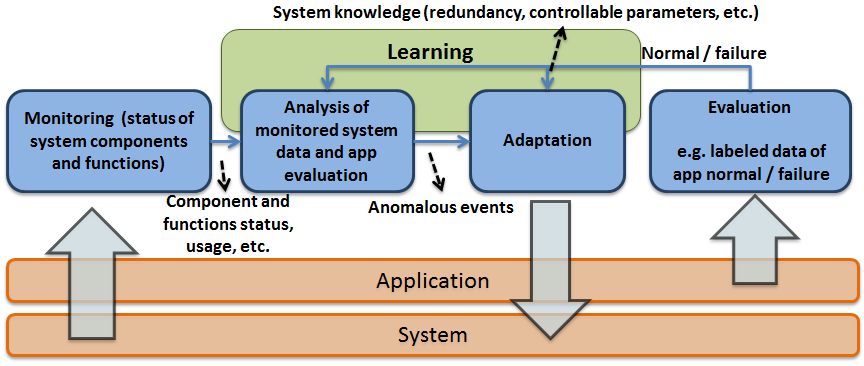}
	\caption{Proactive dependability framework}
	\label{fig:process}
\end{figure}

The proactive dependability framework consists of four mechanisms: \emph{monitoring}, \emph{analysis}, \emph{adaptation}, and \emph{evaluation} (see Figure~\ref{fig:process}). The monitoring mechanism is responsible for monitoring the underlying managed system, and providing the current status of system resources $r_{j}\in R, j=\{1,2,...,m\}$ and functions $f_{k}\in F, k=\{1,2,...,n\}$ to the analysis mechanism at discrete time $t$, where time $t$ refers to a monitoring instance. System resources and functions are monitored by Observable Parameter (OPs) $v_{p}\in V, p=\{1,2,...,m\}$, and $w_{q}\in W, q=\{1,2,...,n\}$ respectively. We need to specify what can be and should be monitored and their normal and failure states' boundaries. For example, RSSI provides information about the connectivity of a device, and battery level provides the remaining energy of a device. The monitoring mechanism $O_{t_{i}}:R\times F\rightarrow V\times W$ for each
time slot $t_{i}\in \lbrack t_{0},t_{1},...,t_{c}]$ is defined as $O_{t_{i}}(r_{1},r_{2},...,r_{m};f_{1},f_{2},...,f_{n})=(v_{1}^{i},v_{2}^{i},...,v_{m}^{i};w_{1}^{i},w_{2}^{i},...,w_{n}^{i})$
where the output is $(m+n)$ dimensional vector.

The analysis mechanism is responsible to analyze the collected data to identify various anomalous events, and predict faults. First, the system needs to learn a threshold $\tau$ for each $v_{p}\in V$ and $w_{q}\in W$, which denotes the boundary between the system $S_{N}$ and $S_{F}$ states. A prediction model is used to predict faults, whenever the value of a system $r_{j}\in R$ or $f_{k}\in F$ corresponding to $v_{p}\in V$ or $w_{q}\in W$ is beyond the threshold $\tau$ (e.g., defined by thresholds of battery level, RSSI level). The outcome of a prediction model is defined by the tuple $(S,F_{type},t_{PF}^{min},t_{PF}^{max})$, where $S$ refers to the system resource or function which has expected fault, $F_{type}$ refers to the type of the predicted fault, and [$t_{PF}^{min}$, $t_{PF}^{max}$] refers to the interval in which the fault is expected.

Adaptation encompasses the mechanisms needed to prevent the predicted faults to keep the system in its normal operation state by autonomously adjusting Controllable Parameters (CPs). The evaluation mechanism checks performance against a set of application requirements to see whether an application is achieving its goals or not. The evaluation mechanism labels the application states with one of two states, namely, normal or failure.

\section{Case Study}
The physical low-level infrastructure of SE applications is based on a wireless sensor network (WSN) platform~\cite{warriach:14b}. First, we investigate two important features of a WSN platform, namely, its ability to communicate and its operational lifetime (battery). The operational lifetime of a WSN node is directly related to its energy source and energy consumption. We identified a number of OPs (RSSI, number of direct neighbors, number of received messages, and current energy level) that can be used to monitor the current status of these features, and to assess the current operational state. Similarly, we identified CPs to control dynamically these features (TX power, RX sensitivity, number of TX and RX schedules, round time, sleep/wake intervals, and active local resources) to keep the system in its normal state. An experimental setup is developed to monitor the status of a system function (communication) and resource (battery) by extending the MyriaNed WSN platform with a monitoring mechanism. This platform has been used for various SE applications, e.g., health care, smart homes, environmental monitoring, and intelligent lighting.

\section{Future Work}
The key issue that we are to address is, how to learn the normal and failure states' boundaries of the system resources and functions at runtime. Another key issue is to develop prediction models for system resources and functions. Further, we need to develop adaptation policies to dynamically prevent the predicted faults using the system knowledge e.g., CPs state boundaries, redundancy, etc. Finally, the most important step is to measure the performance of each adaptation policy with respect to the application availability and reliability, efficient resource management, and the manual interventions from system perspective.


\bibliographystyle{IEEEtran}
\bibliography{paper-com}




\end{document}